\documentclass[11pt]{article}
\usepackage{geometry}                
\usepackage{graphicx}
\usepackage{amssymb}
\usepackage{amsmath}
\usepackage{epstopdf}
\usepackage{authblk}
\usepackage{natbib}
\usepackage{tikz}
\usetikzlibrary {arrows.meta}
\usepackage{amsthm}
\usepackage{multirow}
\usepackage{float}

\usepackage[all]{xy}
\usepackage[plain]{algorithm2e}
\newcommand{\T}{\mathrm{T}}

\usepackage{algorithm2e}
\RestyleAlgo{ruled}
\SetKwComment{Comment}{/* }{ */}

\usepackage{array,colortbl,xcolor}

\title{WENDY: Covariance Dynamics Based Gene Regulatory Network Inference}
\author[1,*]{Yue Wang}
\author[2,3]{Peng Zheng}
\author[4,5,6,7]{Yu-Chen Cheng}
\author[8]{Zikun Wang}
\author[9]{Aleksandr Aravkin}

\affil[1]{Irving Institute for Cancer Dynamics and Department of
  Statistics, Columbia University, New York, NY 10027, USA}
\affil[2]{Institute for Health Metrics and Evaluation, Seattle, WA 98195, USA}
\affil[3]{Department of Health Metrics Sciences, University of Washington, Seattle, WA 98195, USA}
\affil[4]{Department of Data Science, Dana-Farber Cancer Institute, Boston, MA 02215, USA}
\affil[5]{Department of Biostatistics, Harvard T.H. Chan School of Public Health, Boston, MA 02115, USA}
\affil[6]{Center for Cancer Evolution, Dana-Farber Cancer Institute, Boston, MA 02215, USA}
\affil[7]{Department of Stem Cell and Regenerative Biology, Harvard University, Cambridge, MA 02138, USA}
\affil[8]{Laboratory of Genetics, The Rockefeller University, New York, NY 10065, USA}
\affil[9]{Department of Applied Mathematics, University of Washington, Seattle, WA 98195, USA}
\affil[*]{yw4241@columbia.edu}
\date{}                                           

\begin{document}
\maketitle

\begin{abstract}
Determining gene regulatory network (GRN) structure is a central problem in biology, with a variety of inference methods available for different types of data. For a widely prevalent and challenging use case, namely single-cell gene expression data measured after intervention at multiple time points with unknown joint distributions, there is only one known specifically developed method, which does not fully utilize the rich information contained in this data type. We develop an inference method for the GRN in this case, netWork infErence by covariaNce DYnamics, dubbed WENDY. The core idea of WENDY is to model the dynamics of the covariance matrix, and solve this dynamics as an optimization problem to determine the regulatory relationships. To evaluate its effectiveness, we compare WENDY with other inference methods using synthetic data and experimental data. Our results demonstrate that WENDY performs well across different data sets. 
\end{abstract}

\section{Introduction}
In general, a gene is transcribed into mRNA and then translated into proteins. This process, known as gene expression, commonly employs mRNA count or protein count to denote the expression level. In addition to directly changing cell phenotypes \citep{qian2020counting,cheng2023reconstruction}, influencing extracellular processes \citep{axelrod2023drosophila}, or even manipulating macroscopic neurological circuitry \citep{li2021chronic,vijayan2022internal}, certain proteins can affect the transcription of other genes (mutual regulation) or their own corresponding genes (autoregulation). Genes and their regulatory relationships form a gene regulatory network (GRN). 

Determining the GRN structure is a central problem in biology, as it reveals how a living organism is maintained \citep{axelrod3895317role}, and provides control of essential biological processes \citep{wang2020identification}, especially treating cancer \citep{mcdonald2023computational,cheng2023mathematical}. However, directly establishing the GRN using traditional technologies is extremely difficult since they cannot measure the expression levels of many genes within the same cell. Instead, numerous methods have been developed to infer the GRN structure from gene expression data. Particularly, recent advancements in single-cell RNA-sequencing technologies have made it possible to profile the whole transcriptome of single cells at large-scale. However, single-cell RNA-seq can only measure one time point because cells have to be killed during the experimental process, making it challenging to study gene regulation relationships that require multiple observations over time.

In this paper, we focus on a specific data type arising from the following setup: First, implement an intervention that affects gene expression (e.g., drugs). Then measure the expression level (generally mRNA count) of $n$ genes for different single cells at multiple time points, and select the data from time points where the expression has not yet reached a stationary state. Since gene expression at the single-cell level is stochastic, for each time point, we obtain many samples of an $n$-dimensional random vector. However, since we need to kill a cell before measuring its gene expression levels, one cell can only be measured once. Thus, we measure different cells at different time points, and we do not have a joint distribution for gene expression at different time points. Although this approach has become common in recent experimental research \citep{chakraborty2021gene}, and it provides more informative data compared to most other approaches, to our knowledge, there is only one inference method developed specifically for this data type, SINCERITIES~\citep{papili2018sincerities}. A major limitation of SINCERITIES is that it requires data from at least six time points to perform well. Additionally, for single-cell expression data of $n$ genes over $T$ time points, this method only extracts $n(T-1)$ numbers for further analyses, implying low data utilization efficiency. 
There have been many inference methods for single-cell time series gene expression data, where the joint distribution of expression levels at different time points is known \citep{huynh2018dyngenie3,zheng2019bixgboost}. Since obtaining the joint distribution of gene expression is difficult, such methods are usually not practically applicable. There are also many inference methods for single-cell gene expression data measured at a single time point \citep{basso2005reverse,zhang2012inferring}, or bulk level gene expression data measured at multiple time points after interventions \citep{perrin2003gene,ma2020inference,wang2022chronic}. These data types are more common because of their low cost. Nevertheless, they provide less information compared to the data type we examine in this study, namely, time series data from single-cell gene expression. Therefore, while it is feasible to convert our considered data type into these more common forms and use corresponding inference methods, such transformations result in a significant loss of the rich information inherent in the original dataset.

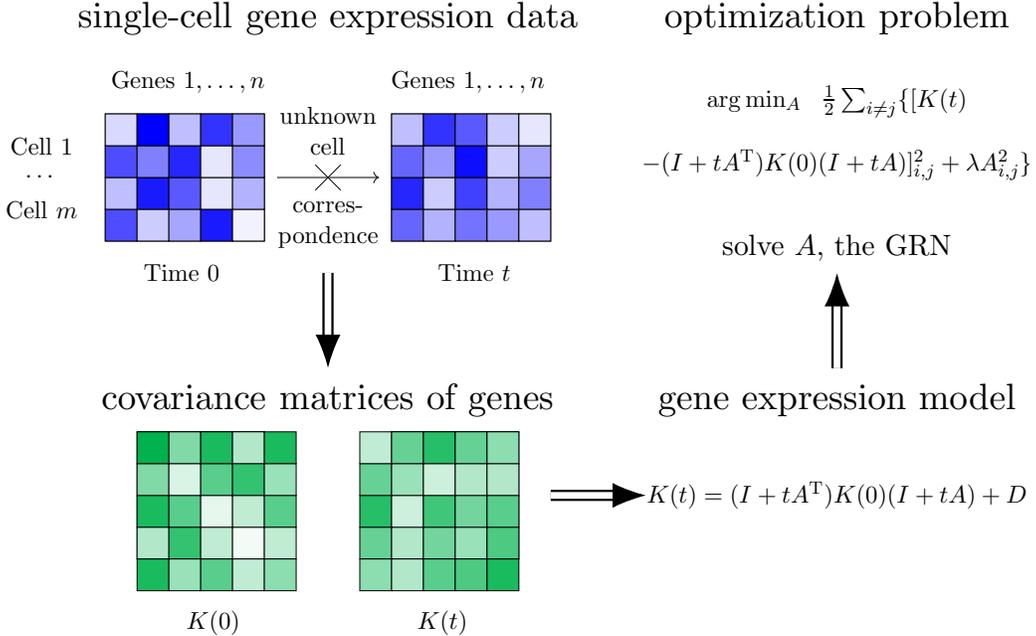
\begin{figure}
    \centering
    \resizebox{5.5 in}{!}{%
    \begin{tikzpicture}
   \draw[black] (-9.5,0) rectangle (-7,2); 
   \draw[black] (-5,0) rectangle (-2.5,2); 
    \draw[->]        (-6.8,1)   -- (-5.2,1);
    \draw       (-6.2,1.2)   -- (-5.8,0.8);
    \draw       (-6.2,0.8)   -- (-5.8,1.2);
    \filldraw[fill=blue!70] (-9.5,0) rectangle (-9,0.5);
    \filldraw[fill=blue!20] (-9,0) rectangle (-8.5,0.5);
    \filldraw[fill=blue!35] (-8.5,0) rectangle (-8,0.5);
    \filldraw[fill=blue!90] (-8,0) rectangle (-7.5,0.5);
    \filldraw[fill=blue!5] (-7.5,0) rectangle (-7,0.5);
    \filldraw[fill=blue!25] (-9.5,0.5) rectangle (-9,1);
    \filldraw[fill=blue!90] (-9,0.5) rectangle (-8.5,1);
    \filldraw[fill=blue!65] (-8.5,0.5) rectangle (-8,1);
    \filldraw[fill=blue!10] (-8,0.5) rectangle (-7.5,1);
    \filldraw[fill=blue!30] (-7.5,0.5) rectangle (-7,1);
    \filldraw[fill=blue!70] (-9.5,1) rectangle (-9,1.5);
    \filldraw[fill=blue!50] (-9,1) rectangle (-8.5,1.5);
    \filldraw[fill=blue!85] (-8.5,1) rectangle (-8,1.5);
    \filldraw[fill=blue!10] (-8,1) rectangle (-7.5,1.5);
    \filldraw[fill=blue!45] (-7.5,1) rectangle (-7,1.5);
    \filldraw[fill=blue!15] (-9.5,1.5) rectangle (-9,2);
    \filldraw[fill=blue!100] (-9,1.5) rectangle (-8.5,2);
    \filldraw[fill=blue!25] (-8.5,1.5) rectangle (-8,2);
    \filldraw[fill=blue!80] (-8,1.5) rectangle (-7.5,2);
    \filldraw[fill=blue!40] (-7.5,1.5) rectangle (-7,2);
    \filldraw[fill=blue!60] (-5,0) rectangle (-4.5,0.5);
    \filldraw[fill=blue!30] (-4.5,0) rectangle (-4,0.5);
    \filldraw[fill=blue!55] (-4,0) rectangle (-3.5,0.5);
    \filldraw[fill=blue!40] (-3.5,0) rectangle (-3,0.5);
    \filldraw[fill=blue!25] (-3,0) rectangle (-2.5,0.5);
    \filldraw[fill=blue!85] (-5,0.5) rectangle (-4.5,1);
    \filldraw[fill=blue!20] (-4.5,0.5) rectangle (-4,1);
    \filldraw[fill=blue!75] (-4,0.5) rectangle (-3.5,1);
    \filldraw[fill=blue!30] (-3.5,0.5) rectangle (-3,1);
    \filldraw[fill=blue!50] (-3,0.5) rectangle (-2.5,1);
    \filldraw[fill=blue!60] (-5,1) rectangle (-4.5,1.5);
    \filldraw[fill=blue!40] (-4.5,1) rectangle (-4,1.5);
    \filldraw[fill=blue!95] (-4,1) rectangle (-3.5,1.5);
    \filldraw[fill=blue!20] (-3.5,1) rectangle (-3,1.5);
    \filldraw[fill=blue!35] (-3,1) rectangle (-2.5,1.5);
    \filldraw[fill=blue!25] (-5,1.5) rectangle (-4.5,2);
    \filldraw[fill=blue!80] (-4.5,1.5) rectangle (-4,2);
    \filldraw[fill=blue!65] (-4,1.5) rectangle (-3.5,2);
    \filldraw[fill=blue!20] (-3.5,1.5) rectangle (-3,2);
    \filldraw[fill=blue!10] (-3,1.5) rectangle (-2.5,2);
\node[scale=1.5] at (-6,3.5) {single-cell gene expression data};
\node at (-10.5,1.5) {Cell 1};
\node at (-6,1.7) {{\begin{tabular}{c} unknown \\ cell \end{tabular}}};
\node at (-6,0.3) {{\begin{tabular}{c} corres- \\ pondence \end{tabular}}};
\node at (-10.5,0.5) {Cell $m$};
\node at (-10.5,1) {$\cdots$};
\node at (-8.2,2.5) {Genes $1,\ldots,n$};
\node at (-3.8,2.5) {Genes $1,\ldots,n$};
\node at (-8.3,-0.5) {Time $0$};
\node at (-3.7,-0.5) {Time $t$};
\draw [line width=1pt, double distance=3pt,
             arrows = {-Latex[length=0pt 3 0]}]        (-6,-0.5)   -- (-6,-2);
\draw[black] (-9,-5) rectangle (-7,-3); 
\draw[black] (-5,-5) rectangle (-3,-3); 
\node[scale=1.5] at (-6,-2.5) {covariance matrices of genes};
    \filldraw[fill=blue!30!green!30] (-9,-5) rectangle (-8.5,-4.5);
    \filldraw[fill=blue!30!green!80] (-8.5,-5) rectangle (-8,-4.5);
    \filldraw[fill=blue!30!green!25] (-8,-5) rectangle (-7.5,-4.5);
    \filldraw[fill=blue!30!green!5] (-7.5,-5) rectangle (-7,-4.5);
    \filldraw[fill=blue!30!green!25] (-7,-5) rectangle (-6.5,-4.5);
    
    \filldraw[fill=blue!30!green!90] (-9,-4.5) rectangle (-8.5,-4);
    \filldraw[fill=blue!30!green!65] (-8.5,-4.5) rectangle (-8,-4);
    \filldraw[fill=blue!30!green!10] (-8,-4.5) rectangle (-7.5,-4);
    \filldraw[fill=blue!30!green!25] (-7.5,-4.5) rectangle (-7,-4);
    \filldraw[fill=blue!30!green!65] (-7,-4.5) rectangle (-6.5,-4);
    
    \filldraw[fill=blue!30!green!50] (-9,-4) rectangle (-8.5,-3.5);
    \filldraw[fill=blue!30!green!15] (-8.5,-4) rectangle (-8,-3.5);
    \filldraw[fill=blue!30!green!65] (-8,-4) rectangle (-7.5,-3.5);
    \filldraw[fill=blue!30!green!80] (-7.5,-4) rectangle (-7,-3.5);
    \filldraw[fill=blue!30!green!40] (-7,-4) rectangle (-6.5,-3.5);
    
    \filldraw[fill=blue!30!green!100] (-9,-3.5) rectangle (-8.5,-3);
    \filldraw[fill=blue!30!green!50] (-8.5,-3.5) rectangle (-8,-3);
    \filldraw[fill=blue!30!green!90] (-8,-3.5) rectangle (-7.5,-3);
    \filldraw[fill=blue!30!green!30] (-7.5,-3.5) rectangle (-7,-3);    
    \filldraw[fill=blue!30!green!90] (-7,-3.5) rectangle (-6.5,-3);

    \filldraw[fill=blue!30!green!90] (-9,-5.5) rectangle (-8.5,-5);
    \filldraw[fill=blue!30!green!40] (-8.5,-5.5) rectangle (-8,-5);
    \filldraw[fill=blue!30!green!65] (-8,-5.5) rectangle (-7.5,-5);
    \filldraw[fill=blue!30!green!25] (-7.5,-5.5) rectangle (-7,-5);
    \filldraw[fill=blue!30!green!45] (-7,-5.5) rectangle (-6.5,-5);

    \filldraw[fill=blue!30!green!60] (-5.5,-5) rectangle (-5,-4.5);
    \filldraw[fill=blue!30!green!30] (-5,-5) rectangle (-4.5,-4.5);
    \filldraw[fill=blue!30!green!55] (-4.5,-5) rectangle (-4,-4.5);
    \filldraw[fill=blue!30!green!40] (-4,-5) rectangle (-3.5,-4.5);    
    \filldraw[fill=blue!30!green!70] (-3.5,-5) rectangle (-3,-4.5); 
    \filldraw[fill=blue!30!green!85] (-5.5,-4.5) rectangle (-5,-4);
    \filldraw[fill=blue!30!green!20] (-5,-4.5) rectangle (-4.5,-4);
    \filldraw[fill=blue!30!green!75] (-4.5,-4.5) rectangle (-4,-4);
    \filldraw[fill=blue!30!green!55] (-4,-4.5) rectangle (-3.5,-4); 
    \filldraw[fill=blue!30!green!65] (-3.5,-4.5) rectangle (-3,-4); 
    \filldraw[fill=blue!30!green!60] (-5.5,-4) rectangle (-5,-3.5);
    \filldraw[fill=blue!30!green!40] (-5,-4) rectangle (-4.5,-3.5);
    \filldraw[fill=blue!30!green!20] (-4.5,-4) rectangle (-4,-3.5);
    \filldraw[fill=blue!30!green!30] (-4,-4) rectangle (-3.5,-3.5); 
    
    \filldraw[fill=blue!30!green!30] (-3.5,-4) rectangle (-3,-3.5); 
    \filldraw[fill=blue!30!green!25] (-5.5,-3.5) rectangle (-5,-3);
    \filldraw[fill=blue!30!green!60] (-5,-3.5) rectangle (-4.5,-3);
    \filldraw[fill=blue!30!green!85] (-4.5,-3.5) rectangle (-4,-3);
    \filldraw[fill=blue!30!green!60] (-4,-3.5) rectangle (-3.5,-3);
    \filldraw[fill=blue!30!green!45] (-3.5,-3.5) rectangle (-3,-3);
    
    \filldraw[fill=blue!30!green!45] (-5.5,-5.5) rectangle (-5,-5);
    \filldraw[fill=blue!30!green!30] (-5,-5.5) rectangle (-4.5,-5);
    \filldraw[fill=blue!30!green!65] (-4.5,-5.5) rectangle (-4,-5);
    \filldraw[fill=blue!30!green!70] (-4,-5.5) rectangle (-3.5,-5);
    \filldraw[fill=blue!30!green!90] (-3.5,-5.5) rectangle (-3,-5);
\node at (-7.8,-6) {$K(0)$};
\node at (-4.2,-6) {$K(t)$};
\draw [line width=1pt, double distance=3pt,
             arrows = {-Latex[length=0pt 3 0]}]        (-2.5,-4)   -- (-1,-4);
\draw [line width=1pt, double distance=3pt,
             arrows = {-Latex[length=0pt 3 0]}]        (2,-2)   -- (2,-0.5);
\node[scale=1.5] at (2,-2.5) {gene expression model};
\node at (2,-4) {$K(t)=(I+tA^\T)K(0)(I+tA)+D$};
\node[scale=1.5] at (2,3.5) {optimization problem};
\node at (2,2.2) {$\arg\min_{A}\ \ \frac{1}{2}\sum_{i\ne j} \{[K(t)$};
\node at (2,1.2) {$-(I+tA^\T)K(0)(I+tA)]^2_{i,j}+\lambda A^2_{i,j}\}$};
\node[scale=1.2] at (2,-0.1) {solve $A$, the GRN};
\end{tikzpicture}
}
    \caption{Workflow of the WENDY method. Given single-cell level gene expression data at two time points, where the joint distribution (cell correspondence) between two time points is unknown, first calculate the covariance matrix of gene expression for each time point. Then use the mathematical gene expression model to derive the equation of covariance matrices. Last, transform this into an optimization problem and solve the GRN numerically.}
    \label{flow}
\end{figure}

In this paper, we introduce an algorithm named NetWork infErence by covariaNce DYnamics (WENDY), designed to connect single-cell gene expression data at different time points, even in the absence of knowledge about the joint distribution. The core idea behind WENDY is to compute the covariance matrices of gene expression levels at two time points and model the evolution of these covariance matrices over time. To infer the GRN, we formulate a non-convex optimization problem based on the dynamics of covariance matrices and derive a numerical solution. For a visual representation of WENDY's workflow, refer to Figure~\ref{flow}.

One of WENDY's key advantages is its requirement of only two time points worth of data. This feature is particularly valuable in scenarios where intervention and / or measurement ultimately result in cell death, precluding measurements at additional time points. However, if data from more time points are available, WENDY can still be applied to each pair of neighboring time points to detect potential rapid changes in the GRN during the experiment. Furthermore, for single cell expression data comprising $n$ genes across $T$ time points, WENDY extracts $(0.5n^2+0.5n)T$ numbers for further analyses, indicating significantly higher data utilization efficiency.

The paper proceeds as follows. In Section~\ref{sec2}, we present a classification framework for gene expression data and review existing GRN inference methods. Section~\ref{sec3} details the WENDY method, including the mathematical gene expression model and the approach to solving the dynamics of this model. In Section~\ref{sec4}, we evaluate WENDY and other GRN inference methods using synthetic data to compare their performance. In Section~\ref{sec5}, we evaluate WENDY and other GRN inference methods using experimental data to compare their performance. Finally, we conclude with discussions in Section~\ref{sec6}.

\section{Data classification and literature review}
\label{sec2}
\subsection{A framework for data classification}
There are different types of gene expression data that can be used to infer the GRN structure. Different data types correspond to different inference methods. We first present a framework for classifying related data types, modified from the framework by \cite{wang2022inference}. See Table~\ref{table1} for this classification framework. There are different dimensions to classify data types. 

(1) We can measure the gene expression levels when the dynamics of gene expression is stationary (invariant along time), or we can add an intervention to drive the dynamics of gene expression away from stationarity, and measure the gene expression levels when they gradually return to the (possibly new) stationary state. For the intervention, we consider general interventions such as adding drugs (we cannot control which genes are affected) and specific interventions such as gene knockdown and gene knockout (we can select any genes to affect). Considering our capability to measure gene expression levels pre- and post-intervention, a specific intervention yields more informative data than a general one. Moreover, scenarios with intervention are richer in information compared to those without, where only stationary expression levels are observed.

(2) We can measure the average expression levels of many cells (bulk level) or measure the expression levels for each single cell (single-cell level). On single-cell level, gene expression is essentially stochastic, and we shall obtain a random variable for the expression level of each gene. On bulk level, the stochasticity is averaged out, and we should obtain a deterministic value for the expression level of each gene. Single-cell level measurement is more informative than bulk level measurement. 

In practice, repeating the same bulk level measurement can still lead to different values, making some researchers regard such data as stochastic and apply inference methods designed for single-cell data \citep{basso2005reverse}. Nevertheless, at bulk level, randomness from single cells is averaged out, and the different values from bulk level measurement can only come from systematic differences, such as different cell phenotypes or different environmental factors. Such unobserved systematic differences can affect multiple genes and make them correlated, although these genes might not have direct regulatory relations. Therefore, we do not consider bulk level data that have different values for the same measurement.

(3) We can measure expression levels at one time point or multiple time points. When we measure expression level at multiple time points, one essential issue is whether we can measure the same cell multiple times. For bulk level data, this does not matter, as the data are deterministic, and whether the cells at $t+1$ are the same as the cells at $t$ should not make a difference. However, for single-cell level data, since the measured levels are stochastic, there is an essential difference. Denote the single-cell expression level of a gene at time $t$ as $X(t)$. If the same cell can be measured multiple times, then we have the joint probability distribution of a time series, $\mathbb{P}[X(0)=x_0,X(1)=x_1,X(2)=x_2,\ldots]$. Otherwise, we only have marginal probability distributions for each time point, $\mathbb{P}[X(0)=x_0]$, $\mathbb{P}[X(1)=x_1]$, $\mathbb{P}[X(2)=x_2]$, $\ldots$, but not the correspondence between time points, and certain quantities cannot be calculated, such as the correlation coefficients of expression levels at two time points. Time series data types are more informative than one-time data types, and joint distribution is more informative than marginal distributions.


In practice, measuring the expression levels of many genes is destructive, and we cannot measure the same cell more than once. If we only want to measure the expression level of a single gene, there are some techniques (fluorescent proteins \citep{wu2011modern}, etc.) that can measure the same cell multiple times. Another approach is to measure the amount of spliced and unspliced mRNAs, which provides both the current expression level and an approximation of its time derivative (RNA velocity \citep{la2018rna}). This approach provides two measurements of the same cell, and some inference methods for time series data can be applied.

	\begin{table}[ht]
		\begin{tabular}{|l|ll|lll|}
			\hline
			\multirow{3}{*}{}                                                   & \multicolumn{2}{l|}{One-time}                                                                                        & \multicolumn{3}{l|}{Time series}                                                                                                                                                                     \\ \cline{2-6} 
			& \multicolumn{1}{l|}{\multirow{2}{*}{Bulk}} & \multirow{2}{*}{\begin{tabular}[c]{@{}l@{}}Single-\\ cell\end{tabular}} & \multicolumn{1}{l|}{\multirow{2}{*}{Bulk}} & \multicolumn{2}{l|}{Single-cell}                                                                                                                        \\ \cline{5-6} 
			& \multicolumn{1}{l|}{}                      &                                                                         & \multicolumn{1}{l|}{}                      & \multicolumn{1}{l|}{\begin{tabular}[c]{@{}l@{}}Marginal\\ distribution\end{tabular}} & \begin{tabular}[c]{@{}l@{}}Joint\\ distri-\\ bution\end{tabular} \\ \hline
			\begin{tabular}[c]{@{}l@{}}Station-\\ ary\end{tabular}              & \multicolumn{1}{l|}{1: No}                 & 2: {\color{red}Yes}                                                                  & \multicolumn{1}{l|}{3: No}                 & \multicolumn{1}{l|}{4: {\color{blue}{\color{blue}Ditto}}}                                                        & 5: {\color{red}Yes}                                                           \\ \hline
			\begin{tabular}[c]{@{}l@{}}General\\ interven-\\ tion\end{tabular}  & \multicolumn{1}{l|}{6: No}                 & 7: {\color{blue}Ditto}                                                                & \multicolumn{1}{l|}{8: {\color{red}Yes}}                & \multicolumn{1}{l|}{\cellcolor{lightgray} 9: {\color{blue}Ditto}/{\color{red}Yes}}                                                       & 10: {\color{blue}Ditto}                                                        \\ \hline
			\begin{tabular}[c]{@{}l@{}}Specific\\ interven-\\ tion\end{tabular} & \multicolumn{1}{l|}{11: {\color{red}Yes$^*$}}               & 12: {\color{blue}Ditto}                                                               & \multicolumn{1}{l|}{13: {\color{blue}Ditto}}             & \multicolumn{1}{l|}{14: {\color{blue}Ditto}}                                                       & 15: {\color{blue}Ditto}                                                        \\ \hline
		\end{tabular}
		\caption{Classification of data types regarding GRN inference, modified from the framework by \cite{wang2022inference}. The data types are classified by different dimensions: (1) The gene expression is at stationary, or is driven away from stationary by an intervention (on general genes that we cannot choose, or specific genes that we can choose); (2) Measure at one time point or multiple time points (time series); (3) Measure the average over many genes (bulk level) or on single-cell level. When measuring at multiple time points on single-cell level, one more dimension is whether we have the joint distribution over different time points. For different data types (scenarios), we study whether the GRN structure can be inferred. For each data type, No means that the GRN structure cannot be inferred. {\color{blue}Ditto} means that there are no specific inference methods, but the GRN structure can be inferred using the same method for a less informative data type. {\color{red}Yes} means specific inference methods exist.  Scenario 11 only has an inference method that also requires the data in Scenario 1. We focus on Scenario 9, which is not as well-studied and has only one specific inference method.}
		\label{table1}
	\end{table}
	
\subsection{Known inference methods for different data types}
Given this classification framework, we can review inference methods for different data types. In this framework, there are 15 data types (scenarios). Some data types do not have enough information that can be used to infer the GRN structure. Some data types (e.g., Scenario 13) have more information than some other data types (e.g., Scenario 8), but the extra information cannot lead to new GRN inference methods. Thus for such scenarios, we can only use methods for other less informative scenarios. This approach loses a lot of information, and therefore cannot justify the time and money spent to obtain more informative data. Some data types have extra information that allows for inference methods that work for such scenarios but not for less informative scenarios.

For bulk level data types, since we only have a single deterministic value for each gene, it is difficult to obtain the correlation between genes. For Scenarios 1, 3, 6, the GRN structure cannot be inferred. For Scenarios 8 and 13, we can regard the gene expression time series as solution trajectories of an ordinary differential equation (ODE) system. If we assume that the ODE system is linear \citep{perrin2003gene} or has certain nonlinear forms \citep{ma2020inference}, we can discretize the ODE system into an algebraic equation system and use regression to infer the ODE parameters. Here the ODE parameters represent the GRN. We can infer all the edges, including the directions. For Scenario 11, one can add an intervention on each gene and observe which genes (descendants of this gene in the GRN) are also affected. Such ancestor-descendant relationships can be used to partially infer the GRN structure \citep{wang2022inference}. Not all regulatory relationships (edges) can be inferred, but one can infer at least $n-1$ edges for a GRN with $n$ genes. 

For single-cell level one-time data types (Scenarios 2, 7, 12), there have been numerous GRN inference methods for Scenario 2, while Scenarios 7 and 12 generally do not have extra information that supports specific inference methods. For Scenario 2, most inference methods turn the GRN inference problem into a feature selection problem: select genes whose levels can be best used to predict the level of the target gene. Then such selected genes might have regulatory relationships with the target gene. The selection can be made by calculating certain quantities between the target gene and the candidate genes that measure their similarity: covariance \citep{nouri2023comparative}, mutual information \citep{basso2005reverse}, or other information theory quantities \citep{zhang2012inferring}. Besides, one can apply regression \citep{haury2012tigress}, decision tree \citep{huynh2010inferring}, or other machine learning and deep learning \citep{shrivastava2022grnular,zhao2022hybrid,zhong2019survival} methods to directly select out genes that can predict the target gene. Regularization terms (e.g., $L_1$ \citep{gustafsson2005constructing} and $L_2$ \citep{kamimoto2023dissecting} regularizers) can be added to the regression to make the result sparse. Besides the idea of feature selection, another approach is to build probabilistic models (Bayesian network \citep{agostinho2015inference,lee2019scaling}, stochastic differential equation (SDE) \citep{wang2023dictys}, and others \citep{liu2016inference,burdziak2023sckinetics}), and use likelihood to determine the most probable network. Since the number of candidate networks is large, a common solution is to apply Markov chain Monte Carlo (MCMC) to approximate the network probabilities \citep{morrissey2010reverse,agostinho2015inference,aalto2020gene,ghosh2021convergence,zhong2021mallows}. Deterministic models, such as Boolean networks \citep{lim2016btr}, can also be used. There is a well-developed platform that combines different inference methods for Scenario 2 \citep{wen2023applying}. One problem of Scenario 2 is to determine the direction of a regulatory relationship, since if the level of $V_i$ can predict the level of $V_j$, then generally the level of $V_j$ can also predict the level of $V_i$. To solve this problem, one can add specific interventions (Scenario 12) on $V_i$ to see whether $V_j$ is affected. 

For single-cell level time series data types without joint distribution of different time points (Scenarios 4, 9, 14), it is common to treat Scenario 4 as Scenario 2 (treat data at different time points separately), and treat Scenarios 9, 14 as Scenario 8 (average over different cells), and apply corresponding inference methods. The only GRN inference method we know that works specifically for Scenario 9 (but not any less informative scenarios) is SINCERITIES \citep{papili2018sincerities}, which considers the Kolmogorov–Smirnov distance between the distributions of the same gene at two time points, and then applies linear regression, similar to methods for Scenario 8. There are two other regression-based methods, Harissa and CARDAMOM \citep{herbach2023harissa,ventre2021reverse}, that essentially work for Scenario 2, but can partially (and insufficiently) integrate the time information when work for Scenario 9.

For single-cell level time series data types with joint distribution of different time points (Scenarios 5, 10, 15), there have been numerous GRN inference methods for Scenario 5, while Scenarios 10 and 15 generally do not have extra information that support specific inference methods. For Scenario 5, most inference methods are similar to those for Scenario 2, especially those methods on regression \citep{zhang2021inference}, tree-based feature selection \citep{huynh2018dyngenie3,zheng2019bixgboost}, or more advanced machine learning tools \citep{nauta2019causal,atanackovic2023dyngfn}, although there are also inference methods based on more complicated biological models \citep{huynh2015combining}. One common approach is to model the gene expression by a vector autoregressive model (either linear or nonlinear) \citep{fujita2007time,fujita2007modeling,siggiridou2015granger}, and then use Granger causality to determine whether one gene directly regulates another gene \citep{fujita2010identification,fujita2010granger,nagarajan2010granger,zhang2010modeling}. This approach solves the famous ``correlation does not imply causation'' problem from two aspects. First, when gene $V_i$ and gene $V_j$ are correlated, it is possible that they are not directly regulating each other, such as $V_i\leftarrow V_k \to V_j$ or $V_i\to V_k \to V_j$. In this case, given the values of $V_k$, $V_i$ cannot provide more information for $V_j$, and Granger causality can determine that $V_i$ does not directly regulate $V_j$. Second, when there is directly regulation between $V_i$ and $V_j$, we do not know whether $V_i$ is the cause or the result of $V_j$. Since Granger causality determines whether the past of $V_i$ contains unique information of the future of $V_j$, the direction of regulation is also known, since causality can only travel forward along time. This explains why most methods for Scenario 5 can determine the direction of regulations, different from their analogies for Scenario 2.

Besides treating a more informative data type as a less informative data type, one can also use certain methods to transform a less informative data type into a more informative data type, provided there are certain assumptions about the underlying systems. For instance, from single-cell one-time data (Scenarios 2, 7, 12), one can construct the pseudotime to transform the data into time series data \citep{reid2016pseudotime,street2018slingshot}, and apply corresponding inference methods \citep{deshpande2022network}. 

Different methods need different assumptions regarding gene expression and gene regulation. For instance, some methods need the gene expression dynamics to be linear \citep{perrin2003gene}, and some other methods need the GRN to have no directed cycle \citep{zhang2012inferring}. Besides, different methods have different inference abilities: some methods can determine all edges, including the direction \citep{huynh2018dyngenie3}, while some other methods can only partially determine some edges, and/or cannot determine the edge direction \citep{wang2022inference}.

Most GRN inference methods can only determine regulations between genes, but not autoregulation. Autoregulation inference methods generally need stronger model assumptions \citep{xing2005causal}, more informative data types \citep{feigelman2016analysis}, or only produce partial results \citep{wang2023inference}.

The above discussion only considers the situation of inferring GRN after obtaining all data. Another situation is to design intervention experiments, so that the GRN can be inferred with the minimal cost \citep{cho2016reconstructing}.

Readers may refer to other reviews for more details about GRN inference methods for different scenarios \citep{barbosa2018guide,pratapa2020benchmarking,zhao2021comprehensive,nguyen2021comprehensive,wang2022inference} or for other data types (besides mRNA/protein count) that can help with GRN inference, such as ChIP-seq and ATAC-seq \citep{erbe2022use,badia2023gene}.

\section{Novel GRN inference method}
\label{sec3}
In this section, we present an algorithm of netWork infErence by covariaNce DYnamics (WENDY), that works for Scenario 9, single-cell level time series data without joint distribution of different time points, measured after general interventions. It can determine all regulatory edges including their directions, but not autoregulation. Using extra information such as DNA sequence and transcription factor binding motifs, some regulatory edges can be excluded. Such prior knowledge can be incorporated by WENDY, but we first consider the original problem that any regulatory edge is possible (except autoregulation).

\subsection{Mathematical model of gene expression}
\label{s31}
We start by building a model of gene expression for a single cell. Although various factors can affect gene expression, we only study regulations between genes. Assume there are $n$ genes $V_1,\ldots,V_n$, which form a GRN. We assume that no other genes affect $V_1,\ldots,V_n$, meaning that there is no hidden variable. Denote the expression levels of $V_1,\ldots,V_n$ by $X_1,\ldots,X_n$. A common model for the dynamics of $X_i$ is a (stochastic) differential equation
\begin{equation}
	\frac{\mathrm{d}X_i(t)}{\mathrm{d}t}=f(X_1,\ldots,X_n)+c_{1,i}-c_{2,i}X_i+\mathrm{noise}.
	\label{ne1}
\end{equation}
Here $c_{1,i}$ means the basal synthesis rate of $X_i$, and $c_{2,i}X_i$ means the total degradation rate of $X_i$. The function $f$ that reflects the interaction between genes can take any form with any number of unknown parameters. The noise term is generally Gaussian. We think that the random fluctuation is proportional to $X_i$, and decide to follow \cite{pinna2010knockouts,papili2018sincerities} to set the noise term to be $\sigma_i X_i\mathrm{d}W_i(t)$, where $\sigma_i$ is an unknown constant, and $W_i(t)$ is a standard Brownian motion. 

Since $f$ in Eq.~\ref{ne1} is an arbitrary function, it is impossible to determine its parameters. We need to add restrictions on the form of $f$. Besides single gene regulations $V_j\to V_i$, transcription factors can bind to enhancers, whose signals are transmitted by coactivators (e.g., mediators) to promoters, which can recruit additional transcription factors and regulate the expression of one or many genes \citep{smith2023decoding,kamal2023granie}. Regulation of gene expression by enhancers can occur over long distances, and in some instances, on different chromosomes. Some enhancers can regulate more than 1 gene \citep{bravo2023scenic+}. Therefore, to model the complicated gene expression, at least we need to consider cooperative regulations with two genes, $V_j+V_k\to V_i$. This means that Eq.~\ref{ne1} becomes 
\begin{equation}
	\frac{\mathrm{d}X_i(t)}{\mathrm{d}t}=\sum_{j=1}^n a_{j\to i}g(X_j)+\sum_{j<k} b_{j,k\to i} h(X_j,X_k)+c_{1,i}-c_{2,i}X_i+\sigma_i X_i\mathrm{d}W_i(t).
	\label{ne2}
\end{equation}
Here $g$ and $h$ are known functions (not necessarily linear), and $a_{j\to i},b_{j,k\to i}$ are regulation coefficients. 

To determine the GRN, we need to determine all coefficients $a_{j\to i},b_{j,k\to i}$, whose total number is $n^3/2+n^2/2$ for $n$ genes. For single-cell gene expression data, at each time point, the data set is a matrix of $m$ cells $\times$ $n$ genes. From the data, we can estimate the parameters of the joint distribution of $n$ genes. If this joint distribution is Gaussian, we can obtain $n$ parameters for the mean, and $n(n+1)/2$ parameters for the (co)variance (since the covariance matrix is symmetric), which fully determine the distribution. Even though the joint distribution of $n$ genes might not be Gaussian in reality, the number of parameters that can be estimated from expression data at each time point should be at the level of $n^2/2$. Therefore, to determine $n^3/2+n^2/2$ GRN parameters in Eq.~\ref{ne2}, we need at least $n$ time points. This generally does not hold in reality, since a small GRN of interest might contain tens of genes, but the number of time points for type 9 data is commonly only a few \citep{hayashi2018single,treutlein2016dissecting,chu2016single}. Thus we have to further (over)simplify Eq.~\ref{ne2} and drop cooperative terms:
\begin{equation}
	\frac{\mathrm{d}X_i(t)}{\mathrm{d}t}=\sum_{j=1}^n a_{j\to i}g(X_j)+c_{1,i}-c_{2,i}X_i+\sigma_i X_i\mathrm{d}W_i(t).
	\label{ne3}
\end{equation}
Now there are only $n^2$ GRN parameters, and theoretically, single-cell data at only $2$ time points can provide enough information. For instance, we can calculate the covariance matrices of $X_1,\ldots,X_n$ at time $0$ and time $t$, and determine what $a_{j\to i}$ in Eq.~\ref{ne3} can lead to such covariance matrices. This is an inverse problem of Eq.~\ref{ne3}, which is much harder than the original problem.

In some models, the function $g$ is nonlinear, meaning that Eq.~\ref{ne3} cannot be solved analytically. Thus, it is very difficult even if we only want to solve the inverse problem numerically, especially if we want the solver to be numerically stable. We need to further (over)simplify the model to assume that it is linear:
\begin{equation}
	\frac{\mathrm{d}X_i(t)}{\mathrm{d}t}=\sum_{j=1}^n a_{j, i}X_j+c_i+\sigma_i X_i\mathrm{d}W_i(t).
	\label{ne4}
\end{equation}
Since Eq.~\ref{ne4} is linear, $a_{i\to i}g(X_i)$ (represents autoregulation) and $-c_{2,i}X_i$ (represents degradation) are combined to $a_{i,i}X_i$, while $a_{j, i}=a_{j\to i}$. This means that we cannot determine the existence of autoregulation, even if we can solve $a_{i,i}$. Since the autoregulation of $V_i$ is masked by the natural synthesis and degradation of $V_i$, most GRN inference methods cannot determine the existence of autoregulation.

We combine Eq.~\ref{ne4} for different $V_i$ to obtain
\begin{equation}
	\frac{\mathrm{d}X(t)}{\mathrm{d}t}=X(t)A+c+X(t)\odot\mathrm{d}\sigma W(t).
	\label{ne5}
\end{equation}
Here $X(t)=[X_1(t),\ldots,X_n(t)]$, $c=[c_1,\ldots,c_n]$, and $\odot$ is the entrywise (Hadamard) product: 
\[X(t)\odot\mathrm{d}\sigma W(t)=[X_1(t)\sigma_1\mathrm{d} W_1(t),\ldots,X_n(t)\sigma_n\mathrm{d} W_n(t)].\]

If we directly solve the covariance matrix of $X_1(t),\ldots,X_n(t)$ from Eq.~\ref{ne5}, then it has an integral that hinders the following optimization procedure. Thus we first ignore the noise term in Eq.~\ref{ne5}, and its solution is 
\[X(t)=[X(0)+cA^{-1}]e^{t A}-cA^{-1}.\]
Since $e^{tA}$ is still difficult to handle in the following optimization procedure, we consider the first-order approximation
\begin{equation}
X(t)=X(0)(I+tA)+tc,
	\label{ne6}
\end{equation}
where $I$ is the $n\times n$ identity matrix. Then we add back the integrated noise term, which is then approximately 
\begin{equation}
X(t)=X(0)(I+tA)+tc + X(0) \odot \epsilon(t),
	\label{ne7}
\end{equation}
where $\epsilon(t)=[\epsilon_1(t),\ldots,\epsilon_n(t)]$ is an $n$-dimensional normal random noise with mean $(0,\ldots,0)$ and covariance matrix $G$. Here we assume that $G$ is diagonal, meaning that noise terms $\epsilon_i(t),\epsilon_j(t)$ for different genes are independent, but the diagonal elements of $G$ are unknown. Besides, $\epsilon(t)$ and $X(0)$ are also independent. 

For type 9 data, after adding drugs or other interventions, we measure the expression levels of $n$ genes at time $0$, and the expression levels are $n$ random variables: $X(0)=[X_1(0),\ldots,X_n(0)]$. Then at time $t$, we measure the expression levels of these $n$ genes again to get $n$ random variables: $X(t)=[X_1(t),\ldots,X_n(t)]$. Since we add interventions to drive the system away from stationary, $X(0)$ and $X(t)$ are not identically distributed. However, we do not have the joint distribution of $X(0)$ and $X(t)$, meaning that we do not know which sample of $X(0)$ corresponds to which sample of $X(t)$. From such data, we want to solve $A$, an invertible $n\times n$ matrix that represents the GRN we want: $A_{i,j}>0$ means gene $i$ activates gene $j$; $A_{i,j}<0$ means gene $i$ inhibits gene $j$; and $A_{i,j}=0$ means gene $i$ does not regulate gene $j$ directly. However, as discussed above, $A_{i,i}$ is the combination of degradation and possibly autoregulation of gene $i$, and $A_{i,i}\ne 0$ does not necessarily mean autoregulation of gene $i$. In the next subsection, we will present a numerical method that calculates $A$ in Eq.~\ref{ne7}.

Gene expression has many biochemical subtleties, and Eq.~\ref{ne7} is certainly an oversimplification of Eq.~\ref{ne1}. Here the simplification from Eq.~\ref{ne1} to Eq.~\ref{ne3} is inevitable, since type 9 data generally do not have many different time points, and cannot provide enough information to solve a model with too many unknown parameters (as in Eq.~\ref{ne2}). If we want to learn some knowledge about the GRN from type 9 data, some simplification is necessary. The simplifications from Eq.~\ref{ne3} to Eq.~\ref{ne7} are only for numerical purposes. In Subsection~\ref{s43}, We will see that although our method is derived from Eq.~\ref{ne7}, it has good performance on data generated by Eq.~\ref{eqnl}, which is in the form of Eq.~\ref{ne3}. Therefore, the simplification from Eq.~\ref{ne3} to Eq.~\ref{ne7} does not harm the generalizability of our method. Besides, although our method cannot determine autoregulation $V_i\to V_i$, it can determine feedback loop $V_i\to V_j\to V_i$, which is the fundamental mechanism of some ``autoregulations'' observed in experiments \citep{crews2009transcriptional}.

\subsection{Dynamics of covariance matrix}

Since we do not know the joint distribution of $X(0)$ and $X(t)$, we cannot determine which sample of $X(0)$ corresponds to which sample of $X(t)$, meaning that directly attacking Eq.~\ref{ne7} does not help. Instead, we can assume that some statistical quantities on both sides should be equal. 

If we take expectation on Eq.~\ref{ne7}, it returns to Eq.~\ref{ne6}. For any $A$, we can find the value of $c$ to make Eq.~\ref{ne6} hold. Thus we cannot solve $A$ from Eq.~\ref{ne6}. Instead, we can consider the covariance matrices of $X(0)$ and $X(t)$, denoted as $K(0)$ and $K(t)$. We have
\begin{equation}
	\begin{split}
		K(t)=&\mathbb{E}\{[X(t)^\T-x(t)^\T][X(t)-x(t)]\}\\
		=&\mathbb{E}\{(I+tA^\T)[X(0)^\T-x(0)^\T][X(0)-x(0)](I+tA)\}\\
		&+\mathbb{E}\{(I+tA^\T)[X(0)^\T-x(0)^\T][X(0) \odot \epsilon(t)]\}\\
		&+\mathbb{E}\{[X(0) \odot \epsilon(t)]^\T[X(0)-x(0)](I+tA)\}\\
		&+\mathbb{E}\{[X(0) \odot \epsilon(t)]^\T[X(0) \odot \epsilon(t)]\}\\
		=&(I+tA^\T)\mathbb{E}\{[X(0)^\T-x(0)^\T][X(0)-x(0)]\}(I+tA)\\
		&+\mathbb{E}\{(I+tA^\T)[X(0)^\T-x(0)^\T][X(0) \odot \mathbb{E}\epsilon(t)]\}\\
		&+\mathbb{E}\{[X(0) \odot \mathbb{E} \epsilon(t)]^\T[X(0)-x(0)](I+tA)\}\\
		&+\mathbb{E}[X(0)^\T X(0)] \odot \mathbb{E}[\epsilon(t)^\T \epsilon(t)]\\
		=&(I+tA^\T)K(0)(I+tA)+D.
	\end{split}
	\label{cov}
\end{equation}
Here $D$ is diagonal, with $D_{i,i}=\mathbb{E}[X_i(0)^2]G_{i,i}$.

Given $K(0)$ and $K(t)$, we cannot solve $A$ directly from Eq.~\ref{cov}, even if we set $D=0$. Assume $K(0)$ and $K(t)$ are invertible. As covariance matrices, they are positive-definite and have Cholesky decomposition $K(0)=L_0^\T L_0$ and $K(t)=L_1^\T L_1$ with upper-triangular $L_0$ and $L_1$. Then for any orthonormal matrix $O$ with $O^\T O=I$, $A=(L_0^{-1}OL_1-I)/t$ is a solution of Eq.~\ref{cov} with $D=0$. Thus Eq.~\ref{cov} has infinitely many solutions in this case, and we need to add some conditions to obtain a unique $A$.

\subsection{Optimization formulation for covariance dynamics}

Assume we measure the expression levels of $n$ genes for $m$ single cells. When $m<n$, which is common in reality, if we directly calculate the covariance matrix, it will always be degenerate (non-invertible). Therefore, we need to apply a specific method, called graphical lasso, that estimates the covariance matrix $K$ in this case, where $K$ is invertible, and the inverse of $K$ is sparse \citep{friedman2008sparse}. 

For a $1\times n$-dimensional random vector $N=[N_1,\ldots,N_n]$ with invertible covariance matrix $K$, there is a result that $K^{-1}_{i,j}=0$ if and only if the partial Pearson correlation coefficient satisfies \citep{wasserman2004all} 
\[\rho_{i,j\mid 1,\ldots,i-1,i+1,\ldots,j-1,j+1,\ldots,n}=0.\] Therefore, if $N$ is multivariate normal, then $K^{-1}_{i,j}=0$ if and only if $N_i$ and $N_j$ are independent conditioned on other variables. 

By assuming the expression levels of $n$ genes satisfy a multivariate normal distribution, there is a GRN inference method \citep{nouri2023comparative}: gene $i$ and gene $j$ have a direct regulatory relation (direction unknown) if and only if $K^{-1}_{i,j}\ne 0$. Nevertheless, this method assumes that the gene expression is in stationary.

For the data set we consider, the intervention might change the dynamics, and the gene expression is not in stationary. Therefore, $K^{-1}_{i,j}\ne 0$ might just mean that gene $i$ and gene $j$ has a direct regulatory relation before the intervention, not necessarily implying $A_{i,j}\ne 0$. However, the inverse should be true: if $K(0)^{-1}_{i,j}= 0$ and $K(t)^{-1}_{i,j}= 0$, then gene $i$ and gene $j$ should have no direct regulatory relation, whether before intervention or after intervention. Define $\mathcal{C}=\{(i,j)\mid K(0)^{-1}_{i,j}= 0 \text{ and }K(t)^{-1}_{i,j}= 0\}$. Then $A_{i,j}=0$ if $(i,j)\in \mathcal{C}$. Since $K(0)^{-1}$ and $K(t)^{-1}$ are symmetric, $\mathcal{C}$ is also symmetric: $(i,j)\in \mathcal{C}$ implies $(j,i)\in \mathcal{C}$. Besides, since $K(0)^{-1}$ and $K(t)^{-1}$ are sparse, $\mathcal{C}$ contains most edges.

Certain data, such as DNA sequence and transcription factor binding motifs and ATAC-seq data, can provide prior knowledge that gene $i$ cannot regulate gene $j$, meaning that $A_{i,j}=0$. We denote the set of such forbidden edges as $\mathcal{F}$. Notice that $\mathcal{F}$ might not be symmetric.

From the data, after estimating the covariance matrices, we obtain invertible covariance matrices $K(0)$ and $K(t)$, where $K(0)^{-1}$ and $K(t)^{-1}$ are sparse. Now we have
\begin{equation}
	K(t)-(I+tA^\T)K(0)(I+tA)=D.
	\label{inv}
\end{equation}
Since the diagonal matrix $D$ is unknown, we only want to match off-diagonal elements of $K(t)$ and $(I+tA^\T)K(0)(I+tA)$. Therefore, we need to solve $A$ from Eq.~\ref{inv} regardless of diagonal elements, so that $(I+tA)_{i,j}=0$ whenever $(i,j)\in \mathcal{C}\cup \mathcal{F}$. Under this restriction, there might not be a solution. Instead, we can minimize the matching error and turn it into an optimization problem:
\begin{equation}
    \min_A f_\lambda(A) := \frac{1}{2}\sum_{i\ne j} \{[K(t)-(I+tA^\T)K(0)(I+tA)]_{i,j}\}^2
    +\lambda A^2_{i,j},
    \label{eqo}
\end{equation}
where $A_{i,j}=0$ whenever $(i,j)\in \mathcal{C}\cup \mathcal{F}$ or $i = j$, while $\lambda \ge 0$ is a predetermined constant. The constraints are handled by only optimizing over the nonzero edges, 
and thus~\eqref{eqo} simplifies to a (possibly) regularized nonlinear least squares problem. 
We set $\lambda=0$ in the following simulations, but allow users to adjust $\lambda$ if necessary.

Since $\mathcal{C}\cup \mathcal{F}$ contains most edges, the final $A$ is sparse, which is biologically favorable, since we do not want a very dense GRN. If calculated $A_{i,j}>0$ or $A_{i,j}<0$ ($i\ne j$), we claim that gene $i$ regulates (activates/inhibits) gene $j$. The diagonal elements of $A$ do not provide information about gene regulation, as we cannot distinguish between autoregulation and normal gene expression.

We use the BFGS algorithm to minimize~\eqref{eqo}. The overall algorithm is presented in Algorithm~\ref{alg1}.
For the WENDY method, step 4 of Algorithm~\ref{alg1} is much faster than step 2, since the solver terminates after a constant number of iterations. Graphical lasso has time complexity $\mathcal{O}(n^3)$ \citep{friedman2008sparse}, which does not depend on the cell number $m$. Therefore, the overall time complexity of WENDY method is $\mathcal{O}(n^3)$. In Section~\ref{sec4}, we will see that in practice, the time cost of WENDY increases with $n$, but not $m$.

We present the workflow of WENDY in Algorithm~\ref{alg1}. See 
\begin{verbatim}
https://github.com/zhengp0/genet
\end{verbatim}
and
\begin{verbatim}
https://github.com/YueWangMathbio/WENDY
\end{verbatim}
for the Python implementation of WENDY.

\begin{algorithm}[!htbp]
	\caption{Detailed workflow of WENDY method.}
	\label{alg1}
	\ \\
	\begin{enumerate}
		{	\item \textbf{Input}: 
			
			\quad Single-cell gene expression data at $T=0$ (for $m_1$ cells and $n$ genes) and at $T=t$ (for $m_2$ cells and the same $n$ genes), both measured after general interventions. Prior knowledge of forbidden edges, $\mathcal{F}$
			
			\item \textbf{Call} graphical lasso method to calculate the covariance matrix $K(0)$ for expression data at $T=0$, so that $K(0)$ is invertible, and $K(0)^{-1}$ is sparse. Also calculate the covariance matrix $K(t)$ for expression data at $T=t$

                \item \textbf{Construct} $\mathcal{C}=\{(i,j)\mid K(0)^{-1}_{i,j}= 0 \text{ and }K(t)^{-1}_{i,j}= 0\}$. 
                
			
			\item \textbf{Call} BFGS solver for the optimization problem 
   \begin{equation*}
       \begin{split}
           \arg\min_{A}\ \ &\frac{1}{2}\sum_{i\ne j} \{[K(t)-(I+tA^\T)K(0)(I+tA)]^2_{i,j}+\lambda A^2_{i,j}\}\\
       \end{split}
   \end{equation*}

                with constraints $A_{i,j}=0$, $\forall (i,j)\in \mathcal{C}\cup \mathcal{F}$ 
			
			\item\textbf{Output}: 
			
			\quad The GRN matrix $A$. $A_{i,j}>0$ or $A_{i,j}<0$ means gene $i$ activates/inhibits gene $j$. However, $A_{i,i}\ne 0$ does not necessarily mean autoregulation of gene $i$
			
		}
	\end{enumerate}
\end{algorithm}

\subsection{Theoretical comparison with other methods}
To infer GRN structure in Scenario 9, besides WENDY, we have four other options: (I) apply SINCERITIES method; (II) calculate the average gene expression levels over all cells to transform the data into Scenario 8, and apply corresponding methods; (III) only consider the data at one time point, which is Scenario 2, and apply corresponding methods; (IV) treat the data at each time point as Scenario 2, and apply corresponding methods, but calibrate the inferred GRN by the GRN inferred for the previous time point, similar to Harissa and CARDAMOM.

(I, II) For an expression level data set with $n$ genes at $T$ time points, SINCERITIES only extracts $n(T-1)$ values, and then applies linear regression. If we transform the data into Scenario 8, we can only obtain $nT$ values before fitting to an ODE system. If we abandon data at other time points to switch to Scenario 2, we lose the temporal information. In comparison, WENDY extracts $n^2T/2+nT/2$ values from such data before proceeding to the next step. Therefore, WENDY can extract more information from data. SINCERITIES requires at least $T=4$ time points to work normally (with $3$ time points, SINCERITIES always produces an all-zero matrix), and requires at least $T=6$ time points to perform well. Methods for Scenario 8 need at least $T=3$ time points to work, and also more time points to work well. In comparison, WENDY only needs $T=2$ time points. This can be explained by information theory: the GRN (without autoregulation) has $n^2-n$ independent values. WENDY uses data at $T=2$ time points to extract $n^2+n$ values; SINCERITIES needs data at $T=n$ time points to extract $n^2-n$ values; methods for Scenario 8 need data at $T=n-1$ time points to extract $n^2-n$ values.

SINCERITIES and methods for Scenario 8 require more time points. This not only increases the cost, but also requires that the GRN does not change among such time points. This might not hold in reality, such as in the early development of embryos, where the regulatory strength might change rapidly. Also, when studying the effect of drugs on gene expression, the regulatory effect of drugs will gradually disappear. Instead, since WENDY only needs two time points, when there are data from more time points, we can apply WENDY for each pair of adjacent time points, and study how the GRN evolves along time.

(III) The theoretical foundation of some methods for Scenario 2 requires that the gene expression is at stationary. For Scenario 9, after adding the intervention, we need to wait for the gene expression to return to stationary state. However, in some experiments, the intervention (e.g., adding certain drugs) can drive the gene expression too far away from stationary, which leads to the death of cells before returning to stationary. Therefore, WENDY is a better solution to this transient scenario.

(IV) Harissa and CARDAMOM \citep{herbach2023harissa,ventre2021reverse} have the same idea: for Scenario 2 data, build a nonlinear regression model under complicated pre-processing. Given the single-cell expression data $D_t$ at time $t$, they can calculate the regression error $R(D_t,A_t)$ for any possible GRN $A_t$. Different from (III), they do not directly minimize $R(D_t,A_t)$, but add a calibration term that prevent the current inferred GRN $A_t$ from being too different with the inferred GRN $A_{t-1}$ for the previous time point $t-1$:
\[A_t=\arg\min_{A}R(D_t,A)+||A-A_{t-1}||_1.\]
When $t=1$, set $A_0$ to be the zero matrix. For Scenario 9 data at time points $1,2,\ldots,T$, use the last inferred GRN $A_T$ as the final result. This idea can partially utilize the time information, but most information between two neighboring time points is wasted. For the situation that the GRN does not change along time, this idea relies heavily on data at later time points, and wastes the data at early time points. For the situation that the GRN can change along time, this idea only reflects the GRN at the last time point, but not GRN at earlier time points. Since these two methods are based on a piecewise-deterministic Markov process model, not the stochastic differential equation model that generate the testing data for the next section, we will not test their performance.

\section{Performance evaluation on synthetic data}
\label{sec4}

In this section, we use two synthetic data sets with different settings to test the performance of WENDY and four other GRN inference methods that work for different data types: GENIE3, dynGENIE3, NonlinearODEs, SINCERITIES. WENDY ranks second for the DREAM4 data set, and ranks third for the SINC data set.

\subsection{Methods and measurements}
GENIE3 \citep{huynh2010inferring} works for Scenario 2, single-cell level one-time expression data at stationary. The idea is to infer the level of one gene by the levels of other genes using random forest or extra trees. All edges can be inferred, but sometimes the direction is hard to determine.

dynGENIE3 \citep{huynh2018dyngenie3} is the revised version of GENIE3 that works for time series data, such as Scenario 5 (and 10), single-cell level time series expression data at stationary or after general interventions. The idea is basically the same as GENIE3, but here it infers the level of one gene at a later time by the levels of other genes at an earlier time. Therefore, all edges including the directions can be determined. dynGENIE3 requires the correspondence of cells at different time points. This means that it cannot be applied to Scenario 9 data directly. Therefore, dynGENIE3 is included just for completeness, not as a main comparison target.

NonlinearODEs \citep{ma2020inference} works for Scenario 8, bulk level time series expression data measured after general interventions. The idea is to fit the data to a nonlinear ODE model. It uses XGBoost to determine the importance score of each edge. There is a tunable parameter $\alpha$ that can be chosen manually or automatically, and we use the \emph{from\_data} mode to determine the parameter $\alpha$ automatically. All edges including the directions can be determined.

SINCERITIES \citep{papili2018sincerities} works for Scenario 9, single-cell level time series expression data measured after general interventions, where the joint distribution at different time points is unknown. The idea is to calculate the distance between the distributions of the same gene at two time points, and then applies linear regression. All edges including the directions can be determined.

Our WENDY method also works for Scenario 9. All edges including the directions can be determined. For the data sets in this section, there is no extra biological knowledge about regulations. Therefore, the set of forbidden edges $\mathcal{F}=\emptyset$.

To test a GRN inference method under different circumstances, a common practice is to use synthetic data \citep{papili2018sincerities,ma2020inference,huynh2018dyngenie3}. The data are generally generated by numerically simulating an SDE system. We will test different inference methods on two data sets: DREAM4 and SINC.

Each inference method obtains a calculated GRN matrix $A'$, whose entries take values in $\mathbb{R}$. To generate the synthetic data, there is a true GRN matrix $A$, which can take values $1/0/-1$. To evaluate the inference result $A'$ with the true GRN matrix $A$, we calculate two measurements, AUROC and AUPR \citep{papili2018sincerities}. AUROC is the area under the curve of true positive rate versus false positive rate, and AUPR is the area under the curve of precision versus recall. They evaluate how the inferred GRN fits the true GRN from different perspectives. Since $A'_{i,i}\ne 0$ does not necessarily mean autoregulation, we do not compare the diagonal elements of $A$ and $A'$. 

AUROC and AUPR are originally defined for binary classification problems, where the ground truth can only take two values (e.g., 1 and 0), but the true GRN in our problem can take three values. We need to generalize the definitions of AUROC and AUPR. AUROC has another equivalent definition: for all data point pairs $(i,j)$ and $(p,q)$ with $A_{i,j}\ne A_{p,q}$, AUROC is the proportion of data point pairs that $(A_{i,j}-A'_{i,j})(A_{p,q}-A'_{p,q})>0$. This definition works for our ternary case. Notice that this ternary definition of AUROC is equivalent to calculate the binary AUROC for data points with $A_{i,j}=1$ or $A_{p,q}=0$, and weight it by the number of data point pairs $A_{i,j}=1,A_{p,q}=0$, and repeat this to $A_{i,j}=1,A_{p,q}=-1$ and $A_{i,j}=0,A_{p,q}=-1$. Then AUPR can be defined similarly. See Algorithm~\ref{alg2} for the calculation procedure of such quantities. When the true GRN can only take two values, these ternary definitions degenerate back to the original binary AUROC and AUPR. Besides, these ternary definitions also lead to values between 0 and 1, where 1 means perfect ordering.

\begin{algorithm}[!htbp]
	\caption{Calculation procedure of AUROC and AUPR.}
	\label{alg2}
	\ \\
	\begin{enumerate}
		{	\item \textbf{Input}: true GRN matrix $A$ and calculated GRN matrix $A'$, both with size $n\times n$. Here $A_{i,j}$ can take values $1/0/-1$, and $A'_{i,j}\in\mathbb{R}$
			
			\item \textbf{Setup}: 

   $\mathcal{U}=\{(i,j)\mid A_{i,j}=1,i\ne j\}$, $u=|\mathcal{U}|$

   $\mathcal{Z}=\{(i,j)\mid A_{i,j}=0,i\ne j\}$, $z=|\mathcal{Z}|$

   $\mathcal{D}=\{(i,j)\mid A_{i,j}=-1,i\ne j\}$, $d=|\mathcal{D}|$
			\item \textbf{Calculate} 

   $\mathrm{AUROC}_{1/0}^{\mathrm{binary}}$ and $\mathrm{AUPR}_{1/0}^{\mathrm{binary}}$ for data points in $\mathcal{U}$ and $\mathcal{Z}$

   $\mathrm{AUROC}_{1/-0}^{\mathrm{binary}}$ and $\mathrm{AUPR}_{1/-0}^{\mathrm{binary}}$ for data points in $\mathcal{U}$ and $\mathcal{D}$

   $\mathrm{AUROC}_{0/-1}^{\mathrm{binary}}$ and $\mathrm{AUPR}_{0/-1}^{\mathrm{binary}}$ for data points in $\mathcal{Z}$ and $\mathcal{D}$
			
			\item \textbf{Average} 

   \[\mathrm{AUROC}^{\mathrm{ternary}}=\frac{uz\ \mathrm{AUROC}_{1/0}^{\mathrm{binary}}}{uz+ud+zd}+\frac{ud\ \mathrm{AUROC}_{1/-1}^{\mathrm{binary}}}{uz+ud+zd}+\frac{zd\ \mathrm{AUROC}_{0/-1}^{\mathrm{binary}}}{uz+ud+zd}\]
   \[\mathrm{AUPR}^{\mathrm{ternary}}=\frac{uz\ \mathrm{AUPR}_{1/0}^{\mathrm{binary}}}{uz+ud+zd}+\frac{ud\ \mathrm{AUPR}_{1/-1}^{\mathrm{binary}}}{uz+ud+zd}+\frac{zd\ \mathrm{AUPR}_{0/-1}^{\mathrm{binary}}}{uz+ud+zd}\]

			\item\textbf{Output}: AUROC and AUPR
			
		}
	\end{enumerate}
\end{algorithm}

\subsection{DREAM4 data set}
The most common synthetic data set used to evaluate GRN inference methods is DREAM4 – In Silico Network Challenge \citep{marbach2012wisdom}. It has multiple challenges, each with multiple data types. We will consider the two in silico challenges and use only the time series data set. There are $5$ GRNs with $10$ genes, each accompanied by $5$ stochastic trajectories at $21$ time points. There are also $5$ GRNs with $100$ genes, each accompanied by $10$ stochastic trajectories at $21$ time points. The time points are equally distributed: 0, 50, 100, 150, 200, 250, 300, 350, 400, 450, 500, 550, 600, 650, 700, 750, 800, 850, 900, 950, 1000. Each GRN is represented by a matrix $A$, where $A_{i,j}=1$ or $A_{i,j}=0$, which means that gene $i$ can or cannot regulate gene $j$. Also, $A_{i,i}=0$ for each gene $i$. These two data sets are denoted as DREAM4 (10 genes) and DREAM4 (100 genes). DREAM4 data are generated by GeneNetWeaver software \citep{schaffter2011genenetweaver}, which integrates some unknown SDE system. 

We apply WENDY, GENIE3, dynGENIE3, NonlinearODEs, and SINCERITIES methods to DREAM4 data sets. To test the performance of different methods on DREAM4 data, we compare different settings of the same method and choose the best one. Specifically, we find that for methods that can use data from multiple time points, it is not always good to use data from all $21$ time points. Instead, data from some time points should be abandoned. Here we list the best settings for each method on each data set: 

\noindent (1) For WENDY, we regard DREAM4 data as Scenario 9 data. For DREAM4 data (10 genes), we use the data at $t=450$ and $t=850$ without the cell correspondence between different time points. For DREAM4 data (100 genes), we use the data at $t=300$ and $t=550$ without the cell correspondence between different time points. 

\noindent (2) For SINCERITIES, we regard DREAM4 data as Scenario 9 data by ignoring the cell correspondence between different time points. For DREAM4 data (10 genes), we use the data at $8$ time points: $t=$(0, 50, 100, 150, 200, 250, 300, 350). For DREAM4 data (100 genes), we use the data at $5$ time points: $t=$(750, 800, 850, 900, 950). 

\noindent (3) For NonlinearODEs, we regard DREAM4 data as Scenario 8 data. For DREAM4 data (10 genes), we use the data at $3$ time points: $t=$(300, 350, 400). For DREAM4 data (100 genes), we use the data at $8$ time points: $t=$(200, 250, 300, 350, 400, 450, 500, 550).

\noindent (4) For GENIE3, we regard DREAM4 data as Scenario 2 data by considering only one time point. For DREAM4 data (10 genes), we use the data at $t=$450. For DREAM4 data (100 genes), we use the data at $t=$350.

\noindent (5) For dynGENIE3, we regard DREAM4 data as Scenario 10 data. For DREAM4 data (10 genes), we use the data at $18$ time points: $t=$(0, 50, 100, 150, 200, 250, 300, 350, 400, 450, 500, 550, 600, 650, 700, 750, 800, 850). For DREAM4 data (100 genes), we use the data at $17$ time points: $t=$(0, 50, 100, 150, 200, 250, 300, 350, 400, 450, 500, 550, 600, 650, 700, 750, 800).

See Table~\ref{dream} for the results, where the AUROC and AUPR are averaged over all 5 GRNs in the same data set. We can see that on average, WENDY is slightly better than NonlinearODEs and GENIE3, which are significantly better than SINCERITIES, while dynGENIE3 is significantly better than all other methods. This is not surprising, since dynGENIE3 can utilize the cell correspondence (joint distribution) of different time points, which is not realistic, but contains more information.

Besides, some results are lower than those values reported in corresponding papers \citep{huynh2010inferring,huynh2018dyngenie3,ma2020inference}, since we only use the time series data, not combining with other data types, and we do not manually fine-tune the parameters accordingly.

\begin{table}[ht]
\caption{AUROC and AUPR of different methods on DREAM4 data sets}
\begin{tabular}{lllllll}
&&&&&&\\
     &       & WENDY & \begin{tabular}[c]{@{}l@{}}SINCE-\\ RITIES\end{tabular} & \begin{tabular}[c]{@{}l@{}}Nonline-\\ arODEs\end{tabular} & GENIE3 & \begin{tabular}[c]{@{}l@{}}dynG-\\ ENIE3\end{tabular} \\ \hline
     DREAM4 & AUROC  &0.65 & 0.52&0.64 & 0.64&  0.75   \\
   (10 genes)  & AUPR &0.34 &0.19 &0.32 & 0.27&  0.54   \\ \hline
    DREAM4 & AUROC  & 0.59 & 0.55     &    0.57       &0.64 &  0.74  \\
  (100 genes)   & AUPR & 0.05  &   0.03     &  0.04        & 0.05 & 0.14 \\ \hline
  & Total &1.63&1.29&1.57&1.60&2.17

     \label{dream}
\end{tabular}
\end{table}

One problem of the DREAM4 data is that each GRN only generates $5$ or $10$ stochastic trajectories (each corresponds to a measured cell population). This fits with the mainstream of bulk level data in the early 2010s, when it was difficult to repeat the measurement many times. Nevertheless, with the development of single-cell RNA sequencing, it is easier to obtain single-cell data from thousands of cells. Therefore, we want to generate single-cell expression data over more cells and test the performance of inference methods under different cell numbers.

\subsection{SINC data set}
\label{s43}
\cite{pinna2010knockouts} and \cite{papili2018sincerities} use the following SDE to generate synthetic data:
\begin{equation}
\mathrm{d}X_j(t)=V\left\{\beta \prod_{i=1}^n \left[1+A_{i,j}\frac{X_i(t)}{X_i(t)+1}\right]-\theta X_j(t)\right\}\mathrm{d} t +\sigma X_j(t)\mathrm{d}W_j(t).
\label{eqnl}
\end{equation}
Here $X_i(t)$ is the expression level of gene $i$ at time $t$, and $W_j(t)$ is a standard Brownian motion, independent with other $W_i(\tau)$. $A_{i,j}$ describes the GRN. $V=30$, $\beta=1$, $\theta=0.2$, $\sigma=0.1$. When $t\to\infty$, each $X_i(t)$ will converge to the stationary value and fluctuate slightly around it \citep{yang2021potentials,cheng2021stochastic,cheng2021asymptotic}.

The following equation is the first-order approximation of Eq.~\ref{eqnl} when $X_i(t)$ is small:
\begin{equation}
\mathrm{d}X_j(t)=V\left\{\beta \left[1+ \sum_{i=1}^n A_{i,j}X_i(t)\right]-\theta X_j(t)\right\}\mathrm{d} t +\sigma X_j(t)\mathrm{d}W_j(t),
\label{eql}
\end{equation}
Also, Eq.~\ref{ne7} is the discretization of Eq.~\ref{eql}: $A$, $b_i$, $F_{i,i}$ in Eq.~\ref{ne7} correspond to $V\beta A-V\theta I$, $V\beta$, $\sigma^2 t$ in Eq.~\ref{eql}. These facts provide the theoretical support that WENDY (derived from Eq.~\ref{ne7}) can work for data generated by Eq.~\ref{eqnl}, since we only care about off-diagonal elements of $A$, which represent mutual regulations between different genes. For other gene regulation mechanisms, we can also use first-order approximation and discretization to obtain Eq.~\ref{ne7} or similar forms. Therefore, WENDY should be applicable to different gene regulation mechanisms.

For Eq.~\ref{eqnl}, we simulate the system with the Euler-Maruyama method \citep{kloeden1992stochastic} for time $t\in[0,1]$ with time step $\Delta t=0.01$. This means treating $\mathrm{d}X_j(t)$ as $X_j(t+\Delta t)-X_j(t)$, $\mathrm{d}t$ as $\Delta t$, and $\mathrm{d}W_j(t)$ as a normal random variable $\mathcal{N}(0,\Delta t)$. 

For the GRN matrix $A$, we use the $40$ networks in \cite{papili2018sincerities}: $10$ \emph{Escherichia coli} networks with $10$ genes, $10$ \emph{E. coli} networks with $20$ genes, $10$ \emph{Saccharomyces cerevisiae} (yeast) networks with $10$ genes, $10$ yeast networks with $20$ genes. Each network has $A_{i,j}=1/0/-1$, and $A_{i,i}=0$ for each $i$. 

For each group of data, we run the simulation $m$ times, where $m$ represents the number of cells/trajectories measured in reality. We set $m=10$, $m=30$, and $m=100$ to test the performance of different methods under different $m$. The initial state $X_i(0)$ is independently and uniformly sampled in $[0,1)$. For each network in Papili Gao et al.'s paper, and each value of $m$, we generate $100$ groups of data. The same as Papili Gao et al.'s paper, the simulation finishes at $t=3.0$. We record the expression levels at the following $11$ time points: $t=$(0.0, 0.3, 0.6, 0.9, 1.2, 1.5, 1.8, 2.1, 2.4, 2.7, 3.0). For the data sets generated by Eq.~\ref{eqnl} on GRNs with $10$ or $20$ genes, we name them SINC ($10$ or $20$ genes) data. Since the initial state is generally different from the stationary state, SINC data should be regarded as Scenario 10 data (single-cell level time series data under general interventions, where the joint distribution of different time points is known).

To test the performance of different methods on SINC data, we compare different settings of the same method and choose the best one. For WENDY, we fix the first time point to be 0.0. For SINCERITIES, NonlinearODEs, and dynGENIE3, we consider contiguous time points that start from 0.0. Specifically, we find that for methods that can use data from multiple time points, it is not always good to use data from all $11$ time points. Instead, data from some time points should be abandoned. Here we list the best settings for each method on each data set:

\noindent (1) For WENDY, we regard SINC data as Scenario 9 data. For both SINC data (10 genes)and SINC data (20 genes), we use the data at $t=0.0$ and $t=0.3$. 

\noindent (2) For SINCERITIES, we regard SINC data as Scenario 9 data by ignoring the cell correspondence between different time points. For both SINC data (10 genes) and SINC data (20 genes), we use the data at $4$ time points: $t=$(0.0, 0.3, 0.6, 0.9). 

\noindent (3) For NonlinearODEs, we regard SINC data as Scenario 8 data by taking average of gene expression levels over different cells. For both SINC data (10 genes) and SINC data (20 genes), we use the data at $10$ time points: $t=$(0.0, 0.3, 0.6, 0.9, 1.2, 1.5, 1.8, 2.1, 2.4, 2.7).

\noindent (4) For GENIE3, we regard SINC data as Scenario 2 data by considering only one time point. For both SINC data (10 genes) and SINC data (20 genes), we use the data at $t=0.3$.

\noindent (5) For dynGENIE3, we regard SINC data as Scenario 10 data. For SINC data (10 genes), we use the data at $4$ time points: $t=$(0.0, 0.3, 0.6, 0.9). For SINC data (20 genes), we use the data at $3$ time points: $t=$(0.0, 0.3, 0.6).

We apply WENDY, SINCERITIES, NonlinearODEs, GENIE3, and dynGENIE3 to SINC data sets and compare the corresponding AUROC and AUPR. Each AUROC or AUPR value is averaged over 2000 simulations (20 GRNs, each with 100 groups of data). See Table~\ref{sinc} for the results. We can see that WENDY is better than SINCERITIES and NonlinearODEs, but weaker than GENIE3 and dynGENIE3 (which does not work for Scenario 9 data).

For different values of cell/trajectory number $m$, we can see that most performance metrics increase with $m$, meaning that more cells provide more information.

\begin{table}[ht]
\caption{AUROC and AUPR of different methods on SINC data sets}
\begin{tabular}{lllllll}
&&&&&&\\
     &       & WENDY & \begin{tabular}[c]{@{}l@{}}SINCE-\\ RITIES\end{tabular} & \begin{tabular}[c]{@{}l@{}}Nonline-\\ arODEs\end{tabular} & GENIE3 & \begin{tabular}[c]{@{}l@{}}dynG-\\ ENIE3\end{tabular} \\ \hline
     SINC $m=10$ & AUROC  & 0.51 & 0.53 & 0.49 & 0.60 & 0.56  \\ 
   (10 genes)  & AUPR & 0.33 & 0.32 & 0.34 & 0.38 & 0.36  \\  \hline
SINC $m=10$ & AUROC  & 0.50 & 0.50 & 0.49 & 0.54 & 0.51  \\ 
  (20 genes)   & AUPR & 0.09 & 0.07 & 0.09 & 0.10 & 0.09  \\  \hline
     SINC $m=30$ & AUROC  & 0.52 & 0.53 & 0.49 & 0.64 & 0.59  \\ 
   (10 genes)  & AUPR & 0.33 & 0.32 & 0.34 & 0.40 & 0.37  \\  \hline
SINC $m=30$ & AUROC  & 0.50 & 0.50 & 0.49 & 0.58 & 0.53   \\
  (20 genes)   & AUPR & 0.08 & 0.08 & 0.09 & 0.12 & 0.09  \\  \hline
     SINC $m=100$ & AUROC  & 0.58 & 0.49 & 0.50 & 0.70 & 0.61  \\ 
   (10 genes)  & AUPR & 0.35 & 0.32 & 0.34 & 0.44 & 0.37  \\  \hline
SINC $m=100$ & AUROC  & 0.51 & 0.47 & 0.49 & 0.64 & 0.55  \\
  (20 genes)   & AUPR & 0.09 & 0.08 & 0.09 & 0.14 & 0.10  \\  \hline
& Total &4.39 &4.21 &4.24 &5.28 &4.73

     \label{sinc}
\end{tabular}
\end{table}

We also test the running time of each inference method. For SINC ($10$ genes) and SINC ($20$ genes) data sets, we measure the execution time (averaged over different GRNs) of each algorithm for $m=10$, $m=30$, and $m=100$ cells. See Table~\ref{timet} for the running time (in the form of mean $\pm$ standard deviation) in different settings. We can see that for each algorithm, the time cost increases with the gene number $n$. In addition, WENDY, SINCERITIES and non-linearODEs are insensitive to the cell number $m$, while the time costs of GENIE3 and dynGENIE3 increase significantly with $m$. When $m$ and $n$ are large, WENDY is roughly the same fast as SINCERITIES and NonlinearODEs, but much faster than GENIE3 and dynGENIE3. Therefore, WENDY has a satisfactory speed.

\begin{table}[ht]
\caption{Computational time for different algorithms. We consider the SINC data of $n=10/20$ genes over $m=10$, $m=30$, or $m=100$ cells. For each situation, we use $20$ different GRNs, and present the computational time as mean $\pm$ standard deviation. All time costs are in seconds, and are measured on a desktop with Intel i7-13700 CPU.}
\begin{tabular}{lllllll}
     &       & WENDY & \begin{tabular}[c]{@{}l@{}}SINCE-\\ RITIES\end{tabular} & \begin{tabular}[c]{@{}l@{}}Nonline-\\ arODEs\end{tabular} & GENIE3 & \begin{tabular}[c]{@{}l@{}}dynG-\\ ENIE3\end{tabular} \\ \hline
    SINC & $m=10$  & $0.22 \pm 0.04$ & $0.09 \pm 0.01$ & $0.19 \pm 0.02$ & $4.27 \pm 0.06$ & $7.28 \pm 0.08$     \\
     $n=10$ & $m=30$  & $0.15 \pm 0.03$ & $0.11 \pm 0.01$ & $0.18 \pm 0.02$ & $4.83 \pm 0.05$ & $15.86 \pm 0.17$     \\
   genes & $m=100$ & $0.14 \pm 0.03$ & $0.12 \pm 0.01$ & $0.19 \pm 0.03$ & $7.32 \pm 0.09$ & $51.79 \pm 0.63$     \\ \hline
 SINC& $m=10$  & $0.38 \pm 0.10$ & $0.42 \pm 0.01$ & $0.35 \pm 0.03$ & $8.55 \pm 0.05$ & $16.24 \pm 0.13$     \\
$n=20$& $m=30$  & $0.34 \pm 0.03$ & $0.45 \pm 0.01$ & $0.36 \pm 0.08$ & $9.95 \pm 0.07$ & $38.21 \pm 0.28$     \\
  genes & $m=100$ & $0.27 \pm 0.02$ & $0.64 \pm 0.01$ & $0.34 \pm 0.04$ & $16.32 \pm 0.17$ & $131.89 \pm 0.76$ 

     \label{timet}
\end{tabular}
\end{table}

\section{Performance evaluation on experimental data}
\label{sec5}

Besides synthetic data, there are also some GRNs determined by experiments, so that we can use them as the ground truth for the corresponding gene expression data \citep{hu2007genetic,reimand2010comprehensive,peter2015genomic}. In this section, we use two experimental data sets in Scenario 9 with known GRNs to test the performance of WENDY and three other GRN inference methods that work for different data types: GENIE3, NonlinearODEs, SINCERITIES. Since each cell is only measured once, we do not have the joint distribution of expression levels at different time points, and dynGENIE3 does not apply. WENDY ranks first for the THP-1 data set, and ranks second for the hESC data set.

\subsection{THP-1 data set}
THP-1 data set considers single-cell expression levels of monocytic THP-1 human myeloid leukemia cells, measured at $8$ time points, $t=$ 0, 1, 6, 12, 24, 48, 72, 96h, after applying phorbol-12-myristate-13-acetate \citep{kouno2013temporal}. For each time point, there are 120 cells measured. However, since each cell can be measured only once, we do not know how cells at different time points correspond to each other. Therefore, this data set is in Scenario 9. 

The same as \cite{papili2018sincerities}, we consider the following 20 genes: BCL6, CEBPB, CEBPD, EGR2, FLI1, HOXA10, HOXA13, IRF8, MAFB, MYB, NFATC1, NFE2L1, PPARD, PPARG, PRDM1, RUNX1, SNAI3, TCFL5, TFPT, UHRF1. For such genes in THP-1 cells, there has been a GRN determined by experiments \citep{tomaru2009regulatory,vitezic2010building} that we can use as the ground truth.

To test the performance of different methods on THP-1 data, we compare different settings of the same method and choose the best one. Specifically, we find that for methods that can use data from multiple time points, it is not always good to use data from all $8$ time points. Instead, data from some time points should be abandoned. Here we list the best settings for each method:

\noindent (1) For WENDY, we regard THP-1 data as Scenario 9 data. We use the data at $t=0$ and $t=48$ without the cell correspondence between different time points.  

\noindent (2) For SINCERITIES, we regard THP-1 data as Scenario 9 data. We use the data at $7$ time points: $t=$(1, 6, 12, 24, 48, 72, 96). 

\noindent (3) For NonlinearODEs, we regard THP-1 data as Scenario 8 data by taking average of gene expression levels over different cells. We use the data at $3$ time points: $t=$(0, 1, 6). 

\noindent (4) For GENIE3, we regard THP-1 data as Scenario 2 data by considering only one time point. We use the data at $t=24$.

We apply WENDY, SINCERITIES, NonlinearODEs, and GENIE3 to THP-1 data set and compare the corresponding AUROC and AUPR. See Table~\ref{thp1} for the results. We can see that WENDY is better than all other methods. Here GENIE3 has the worst performance, possibly because it cannot determine the sign (activation/inhibition) of a regulation. The results for SINCERITIES are slightly different from those reported by \cite{papili2018sincerities}, possibly because we do not consider diagonal elements of the GRN in the calculation of AUROC and AUPR.

\begin{table}[ht]
\caption{AUROC and AUPR of different methods on THP-1 data set}
\begin{tabular}{lllll}
&&&&\\
    & WENDY & \begin{tabular}[c]{@{}l@{}}SINCE-\\ RITIES\end{tabular} & \begin{tabular}[c]{@{}l@{}}Nonline-\\ arODEs\end{tabular} & GENIE3  \\ \hline
    AUROC  &0.62 & 0.63&0.55 & 0.42   \\
    AUPR &0.51 &0.43 &0.39 & 0.36 

     \label{thp1}
\end{tabular}
\end{table}

\subsection{hESC data set}
hESC data set considers single-cell expression levels of human embryonic stem cell-derived progenitor cells, measured at $6$ time points, $t=$0, 12, 24, 36, 72, 96h \citep{chu2016single}. For each time point, there are 66--172 cells measured. Such cells are under development (which can be regarded as an intervention), and the regulation mechanism of their gene expression is highly complicated \citep{li2017ground}. The same as THP-1 data set, there is no cell correspondence between different time points, and it is in Scenario 9.

\cite{matsumoto2017scode} calculated the mean expression level of each gene at each time point, and then calculated the variance of the mean expression levels at different time points for each gene. Using this approach, they selected out top 100 highly varying genes, which might actively regulate each other. Given the data of these 100 highly varying genes, we further select out genes that express in at least 95\% of the cells. The reason is that when the expression data have too many $0$s, graphical lasso might fail to converge, which fails WENDY. There are 18 genes selected: POU5F1, AEBP2, MIER1, SMAD2, ZNF652, ZFX, TERF1, SOX11, BBX, ZFP42, TULP4, ZNF471, ARID4B, ZNF483, SHOX, ZNF587, ZFP14, CEBPZ. For such genes, there has been a ground truth GRN determined by experiments \citep{neph2012circuitry,stergachis2014conservation} that we can use as the ground truth. 

Similar to the THP-1 data set, we apply WENDY (use $t=12,24$ data), SINCERITIES (use $t=0, 12, 24, 36, 72$ data), NonlinearODEs (use $t=24, 36, 72, 96$ data), and GENIE3 (use $t=24$ data) to hESC data set and compare the corresponding AUROC and AUPR. See Table~\ref{hESC} for the results. We can see that WENDY has the highest AUROC, but the total performance ranks second, worse than SINCERITIES. 

\begin{table}[ht]
\caption{AUROC and AUPR of different methods on hESC data set}
\begin{tabular}{lllll}
&&&&\\
    & WENDY & \begin{tabular}[c]{@{}l@{}}SINCE-\\ RITIES\end{tabular} & \begin{tabular}[c]{@{}l@{}}Nonline-\\ arODEs\end{tabular} & GENIE3  \\ \hline
    AUROC  &0.70 & 0.67&0.61 & 0.64   \\
    AUPR &0.05 &0.33 &0.06 & 0.05 
     \label{hESC}
\end{tabular}
\end{table}

\section{Discussion}
\label{sec6}

In this paper, we address the GRN inference problem for single-cell time series gene expression data following general interventions, where the joint distribution of different time points is unknown. Although this type of data is common in recent experiments, there are few GRN inference methods that fully utilize the information contained in the data. Therefore, we introduce WENDY, a GRN inference method developed for single-cell gene expression data spanning two time points after an intervention. This method is capable of inferring all mutual regulatory relations, including direction. We test WENDY and other GRN inference methods on two synthetic data sets and two experimental data sets, and the performance of WENDY is satisfactory: it is only weaker than dynGENIE3, which does not work on Scenario 9 data. See Table~\ref{tablesum} for a summary. Besides, the time cost of WENDY is almost as low as the fastest method.

\begin{table}[ht]
\caption{Ranking of five GRN inference methods on four data sets}
\begin{tabular}{llllll}
&&&&\\
    & WENDY & \begin{tabular}[c]{@{}l@{}}SINCE-\\ RITIES\end{tabular} & \begin{tabular}[c]{@{}l@{}}Nonline-\\ arODEs\end{tabular} & GENIE3 & \begin{tabular}[c]{@{}l@{}}dynG-\\ ENIE3\end{tabular} \\ \hline
    DREAM4  & 2nd & 5th&4th & 3rd & 1st \\
    SINC &3rd &5th &4th & 1st & 2nd\\
    THP1 &1st &2nd &3rd & 4th & -\\
    hESC &2nd &1st &4th & 3rd & -
     \label{tablesum}
\end{tabular}
\end{table}

The model in Subsection~\ref{s31} only considers cells of the same type and under the same environment. In reality, cells of a complex organism are under different regulations by environmental factors, especially during development. For instance, during animal morphogenesis, retinoic acid can form a concentration gradient (positional information) \citep{wang2020biological}, and leads cells at different positions to different fates by regulating certain genes \citep{wolpert1989positional}. If the levels of such spatially heterogeneous factors can be measured for each single cell, then one can add them into the GRN inference procedure. Otherwise, they become hidden variables in the gene expression model, and can make the inference much more difficult \citep{lo2015time}. Besides such position-related factors, cell type can affect gene regulation \citep{zhang2023inference}. One should label the cell type during experiments or after experiments using cluster analysis \citep{bocci2022splicejac}. If the GRN inference is directly applied to data from different types of cells, the inference result might be unreliable \citep{wang2023inference}. In this work, we assume that cells are of the same type, and they are under the same environment. An important future direction is to build more realistic gene expression models and develop corresponding GRN inference methods.

Although most gene expression models are similar to Eq.~\ref{ne1} that only considers the (mRNA) levels of $V_1,\ldots,V_n$, some researchers argue that the actual gene expression mechanism should be more complex. In some models \citep{bonnaffoux2019wasabi,ventre2021reverse,herbach2023harissa,ventre2023one}, each gene can switch between ``on'' and ``off'' states, which correspond to different transcription rates. Some proteins can affect the transition rates between ``on'' and ``off'' states, which is the only way of gene mutual-regulation. When one gene is turned on, the number of its mRNA increases quickly, until it is turned off, when the mRNA count starts to decay exponentially. This leads to the mRNA bursts phenomenon \citep{elgart2011connecting}, meaning that the mRNA count does not fluctuate locally, but has global bursts. Since proteins are translated from mRNAs, proteins also have bursts. However, proteins degrade much slower than mRNAs, meaning that mRNA count and protein count do not match exactly. Therefore, in such models, besides the mRNA counts $X_1,\ldots,X_n$, there are also hidden variables for gene states and protein counts. However, more complicated models, generally using piecewise-deterministic Markov processes, are harder to solve, and researchers need to conduct large-scale simplifications. Besides, more complicated models need more accurate data, but the current mRNA measurement techniques are not very sensitive \citep{zhang2021single}. Therefore, we think that it is worthwhile to study differential equation based models.

Similar to most other GRN inference methods, WENDY cannot infer autoregulation. One potential future direction is to develop an autoregulation inference method inspired by the principles of WENDY. For instance, we can choose a reasonable nonlinear gene expression model that allows autoregulation, and study whether autoregulation can make a difference for the dynamics of covariance matrices.

In interventional experiments, it is customary to measure expression levels before intervention as the control group. These control data align with Scenario 2 data, enabling the application of corresponding inference methods. Therefore, a comparison of GRNs before and after intervention can elucidate the effect of the intervention on gene regulation.

As of 2024, single-cell level gene expression measurement remains relatively insensitive. It is common to miss all mRNAs of one gene in a single cell, resulting in experimental data with many zeros. This characteristic may impede graphical lasso, and consequently, WENDY. A potential future direction involves developing more robust inference methods capable of handling data with numerous missing values.

Although WENDY aims for data type 9, other types of data can help to improve the results of WENDY. For instance, one can use a CRISPR gene knockout study to verify regulations inferred by WENDY. Besides, advancements in biotechnology introduce new types of data, which are not in our classification framework, but might be applicable to GRN inference, and can enhance WENDY's capabilities through incorporation.

In our tests, we observed that WENDY performs better for data in earlier time points when dealing with a small number of genes (10 or fewer). Conversely, for data collected long after intervention, where gene expression approaches a new steady state, the inference results are less reliable compared to earlier time points. One possible reason is that after enough time, the covariance matrix approaches its steady state, and the small changes of covariances might be covered by random noise. Therefore, when applying WENDY to experimental data, caution should be exercised with results obtained several days after the intervention.

Theoretically, the GRN matrix $A$ calculated by WENDY is not symmetric, allowing determination of the directions of the regulatory relations ($i \to j$ or $j \to i$). However, our simulations indicate that $A_{i,j}$ and $A_{j,i}$ generally exhibit proximity, resulting in a significant decrease in AUPR. A prospective avenue involves developing a novel solver capable of producing highly asymmetric results.

In our WENDY implementation, we employ the standard graphical lasso algorithm to determine the inverse of the covariance matrix. Other algorithms with similar purposes \citep{vinci2019graph,danaher2014joint} can also be considered for integration into WENDY.

\section*{Acknowledgments}
The authors would like to thank anonymous reviewers for helpful comments. YW would like to thank 
Ulysse Herbach and Joseph Zhou for helpful discussions.

\section*{Data and code availability}
The solver used in WENDY method is in 
\begin{verbatim}
https://github.com/zhengp0/genet
\end{verbatim}

Main function of WENDY method (including a tutorial), and other data and code files used in this paper are in 
\begin{verbatim}
https://github.com/YueWangMathbio/WENDY
\end{verbatim}
\bibliographystyle{apalike}
\bibliography{cov}

\end{document}